# BrainPath: A Biologically-Informed AI Framework for Individualized Aging Brain Generation

**Yifan Li, Javad Sohankar, Ji Luo, Jing Li*, Yi Su***

***Abstract*— The global population is aging rapidly, and aging is a major risk factor for various diseases. It is an important task to predict how each individual's brain will age, as the brain supports many human functions. This capability can greatly facilitate healthcare automation to enable personalized, proactive intervention and efficient healthcare resource allocation. This task, however, is extremely challenging due to the brain's 3D complex anatomy. While successes have been made in natural image generation and brain MRI synthesis, existing methods fall short in generating individualized, anatomically-faithful aging brain trajectories. To address the gaps, we propose BrainPath, a novel AI model that, given a single structural MRI of an individual, generates synthetic, longitudinal MRIs representing that individual's expected brain anatomy as they age. BrainPath introduces three architectural innovations: an age-aware encoder with biologically-grounded supervision, a differential age-conditioned decoder for anatomically-faithful MRI synthesis, and a swap-learning strategy that implicitly separates stable subject-specific anatomy from aging effects. We further design novel biologically-informed loss functions including an age calibration loss and an age and structural perceptual loss to complement the conventional reconstruction loss. This enables the model to capture subtle, temporally-meaningful anatomical changes with aging. We apply BrainPath to two of the largest public aging datasets, and conduct comprehensive, multi-faceted evaluation. Our results demonstrate BrainPath's superior performances in generation accuracy, anatomical fidelity, and cross-dataset generalizability, outperforming competing methods.**

***Note to Practitioners*— As people live longer, doctors and health systems increasingly need AI tools that can predict how an individual's health will change with age. BrainPath is designed to help meet this need by creating a personalized "movie" of how a person's brain is expected to age, using only a single MRI scan as input. Instead of relying on population averages or generic aging patterns, BrainPath learns the unique structural details of each person's brain and then simulates how those structures are likely to evolve over time. This is important because many brain diseases begin with subtle changes that are difficult to detect early. By providing a realistic aging baseline for each person, BrainPath can help clinicians spot abnormal or accelerated changes sooner, guide earlier interventions, and improve long-term monitoring. The tool may also support hospital planning and preventative care by enabling more individualized forecasts rather than one-size-fits-all estimates.**

Mr. Y. Li's work is supported by Tsinghua University No.724B2019, Dr. Su is supported by the Arizona Department of Health Services (ADHS), and the state of Arizona (ADHS Grant No. CTR057001).

This work was performed while Yifan Li was a visiting scholar at GT.

Y. Li is with ISYE, Georgia Institute of Technology and the Department of Industrial Engineering, Tsinghua University. Dr. J. Sohankar, J. Luo and S. Yi are with Banner Alzheimer's Institute. Prof. Jing Li is with ISYE, Georgia Institute of Technology.

Corresponding author: Prof. Jing Li and Dr. Su Yi.
e-mail: jli3175@gatech.edu, Yi.Su@bannerhealth.com.

## I. INTRODUCTION

As the global population of older adults continues to grow, the number of people aged 65 and older is projected to rise from 857 million in 2024 to 1.6 billion by 2050[1]. Aging is the primary risk factor for a wide range of debilitating conditions such as Alzheimer's disease (AD), cancer, and heart disease[2]. These facts create an urgent need for computational tools that can predict each individual's aging trajectory. Such predictions can greatly facilitate healthcare automation to enable personalized, proactive intervention and efficient healthcare resource allocation.

Among the organs affected by aging, predicting the brain's aging trajectory is particularly important. The brain supports cognition, memory, emotion, and motor control. Aging-related changes in brain structure are closely linked to cognitive decline, dementia, and loss of independence. The ability to accurately predict how a person's brain is expected to change over time would provide a foundation for early risk stratification and targeted prevention.

Predicting how the brain ages is an extremely challenging task. The brain is a 3D organ with complex anatomy. To effectively capture its anatomy, structural magnetic resonance imaging (MRI) provides an essential tool. It offers high-resolution measurements of brain structure, and has been widely used in brain research. Ideally, densely sampled longitudinal MRI for each individual would provide the best way to capture individual brain aging trajectory. However, frequent MRI scanning of healthy individuals is impractical due to high cost. Even if such data were available, longitudinal MRI would only observe the trajectory, while predicting the trajectory is critical for early healthcare decision.

Therefore, the objective of our study is to develop a novel AI model that, given a single baseline MRI of an individual and any target age, generates a synthetic MRI that represents that person's expected brain anatomy at the target age, thereby forming an individualized aging trajectory. This research is related to several existing research areas, but none of them suffice.

The advances in natural image generation have led to highly successful models for synthesizing realistic images, such as GANs and diffusion models [3, 4]. However, these models cannot be directly used for our problem due to several reasons:

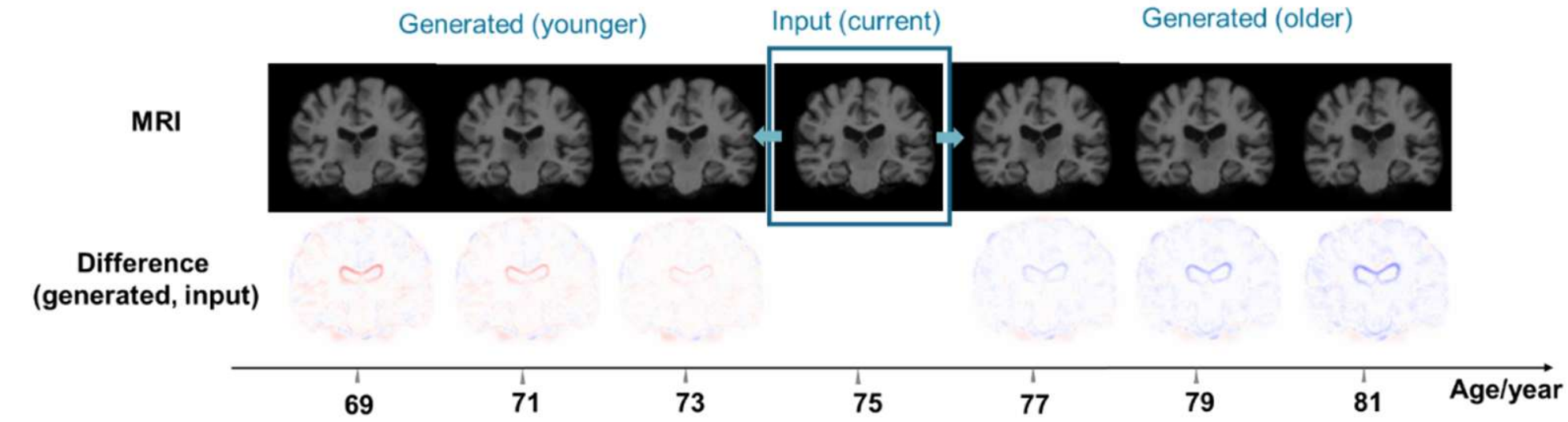

**Fig. 1.** Illustration of BrainPath capability: given the current MRI of an individual, BrainPath can generate future and past MRIs at any given time, revealing individual brain aging trajectory. In the difference maps (second row), red indicate increased intensity (i.e., tissue growth), while blue indicate decreased intensity (i.e., tissue shrinkage).

First, these methods are designed for 2D images, whereas MRI is 3D. Also, brain MRI contains fine-grained anatomical details that are substantially more complex than natural image semantics. Furthermore, MRI data has low signal-to-noise ratio, because age-related anatomical changes are subtle and gradual, whereas image acquisition and preprocessing can introduce noise. This makes aging MRI generation inherently difficult. Additionally, available MRI datasets for normal brain aging contain far fewer samples than natural image datasets. Even when combining multiple large public studies such as Alzheimer's Disease Neuroimaging Initiative (ADNI) [5] and National Alzheimer's Coordinating Center (NACC) [6] , as done in this paper, the total number of cognitively normal subjects and scans is on the order of hundreds to a few thousand, whereas natural image datasets often contain millions of images.

Recent studies have extended image generation techniques to create synthetic brain MRIs. There are two main lines of research. The first one focuses on generating synthetic brain MRIs for data augmentation or disease simulation. These methods aim to increase sample size, enhance image diversity, or create virtual scans with specific pathologies [7, 8]. However, they are not designed to model how an individual's brain changes over time, i.e., they do not preserve subject-specific anatomy with aging. The second line of research is more closely related to our goal and uses generative models to predict future MRIs from past scans. Examples include 2D conditional GANs that "age" brain slices [9, 10], models such as IdenBAT that enforce latent orthogonality between identity and age [11], and longitudinal frameworks that extrapolate a follow-up MRI at fixed time intervals from one or more prior scans [9, 10]. However, the existing methods have several limitations: Some rely on 2D slices, while others require multiple prior scans as input or restrict prediction to predetermined time points. Also, most methods depend on adversarial training, which can be unstable and compromise anatomical fidelity, especially for subtle aging-related changes.

To address the gaps, we introduce BrainPath, a novel 3D model that predicts subject-specific, high-fidelity brain MRIs at any future (or past) timepoints from a single input scan (Fig.1). The contributions of this study are summarized as follows:

- **Novel architecture tailored for aging brain MRI generation**. BrainPath adopts a U-Net–style encoder–decoder framework as the backbone but introduces three key innovations, including (i) an age-aware encoder with a biologically-grounded supervision strategy, (ii) a differential age-conditioned decoder that uses brain-age difference to guide anatomically faithful MRI synthesis, and (iii) a swap-learning mechanism that implicitly separates individual's stable anatomy from aging effects.
- **Novel biologically-informed loss function**. Conventional reconstruction loss in image generation is insufficient for our problem. We introduce two biologically-informed losses: (i) *an age calibration loss* that enforces consistent aging rates within subjects and unbiased brain-age predictions at the population level, and (ii) *an age and structural perceptual loss* that matches feature-level representations and predicted ages between generated and real MRIs. These new losses also ensure our model capture subtle, temporally meaningful anatomical variations that are often obscured by imaging artifacts or suboptimal image preprocessing.
- **Comprehensive multi-dataset evaluation**. We include two of the largest public datasets for aging studies, ADNI and NACC. This allows us to evaluate the true generalizability of our model. Also, we evaluate BrainPath in many aspects, including quantitative and qualitative accuracy, effectiveness in capturing brain aging dynamics, preservation of subject-specific anatomy, and region-specific anatomical fidelity. To our best knowledge, no prior study has performed such a comprehensive, multi-faceted evaluation and utilized more than one dataset.
- **Multi-faceted utilities for healthcare automation**. BrainPath provides a high-fidelity simulation of an individual's expected aging brain trajectory based on their current scan. This will enable early risk identification (e.g., accelerated aging patterns) and proactive, personalized intervention planning. Also, the simulated trajectory establishes an individualized aging baseline of "normal aging" for a person. Deviation between the simulated, normative baseline and a future real MRI can help sensitively detect abnormal changes such as early neurodegeneration or unexpected deterioration.

## II. Literature Review

Our method is related to several existing research areas. Below we review each one and point out the limitations.

### *A. Chronological Age Estimation using MRI*

A prominent line of research uses structural MRI to estimate an individual's chronological age. Various machine learning (ML) and deep learning (DL) methods have been developed. Such research has at least two important utilities: First, accurate estimation demonstrates that MRI contains rich age-related signals. Second, once an MRI-chronological age model is trained on healthy control subjects, the gap between the predicted and true chronological age can be used as an indicator to identify subjects with aging-related diseases such as AD [12]. The ML/DL methods range from early linear models[13], to more sophisticated twin support-vector regression[14], stacked random-forest ensembles[15], and deep learning models like the lightweight simple fully convolutional network (SFCN) [16]. Additional efforts have been made to combine multimodal images [17] and apply transfer learning strategies to adapt pretrained models [18]. Overall, current methods typically report mean absolute errors (MAE) of approximately 3–5 years.

This line of research provides compelling evidence that MRI contains rich age-related information. This helps establish the feasibility of our research in generating MRIs reflective of aging. However, this research area is fundamentally different from what we aim to achieve. Chronological age estimation reduces aging to a single scalar value, which is an extreme simplification of the underlying complex biological process. Such models offer no insight into how the 3D brain anatomy evolves over time.

### *B. General-Purpose Image Generation Models*

Recent advances in natural image generation have led to highly successful models for synthesizing realistic images. These include generative adversarial networks (GANs) [3], which have enabled substantial improvements in image quality through progressively growing architectures [19], style-based modulation [20], and latent disentanglement mechanisms [21]. More recently, diffusion-based models [4] have further advanced controllable image generation through stable training procedures and mechanisms such as classifier-free guidance and cross-attention conditioning [22]. These developments demonstrate that modern generative models can learn structured latent spaces and support meaningful manipulation of semantic attributes.

However, general-purpose image generation methods differ fundamentally from what is required for modeling brain aging. They are designed for 2D natural images with high signal-to-noise ratios and consistent visual semantics. In contrast, 3D brain MRIs contain complex anatomical structures, small-scale aging-related changes, and strict biological constraints that must be preserved. Furthermore, natural-image models do not incorporate mechanisms to ensure anatomical fidelity, longitudinal consistency, or subject-specific structural preservation. As a result, although these generative techniques demonstrate the power of modern image synthesis, they cannot be directly applied to generate 3D MRIs that reflect individualized brain aging trajectories.

### *C. Synthetic MRI Generation*

Recent studies have extended image generation techniques to create synthetic brain MRIs. Most approaches are based on GANs or diffusion models. For instance, a 4D-DANI-Net simulates progressive atrophy sequences to supply virtual follow-ups for dementia studies[23], while other GAN-based methods synthesize images with pathologies such as brain tumors using style-transfer or ensemble strategies [7, 8]. Some diffusion-based approaches also have demonstrated the capability to generate high-resolution 3D brain MRI [24]. Furthermore, conditional diffusion frameworks have been widely adopted to manipulate specific phenotypes[25], where text-guided or attribute-conditioned mechanisms allow for the synthesis of age-related changes or the generation of counterfactual disease states to aid in anomaly detection and data augmentation[26].

However, synthetic brain MRI generators are not designed to model subject-specific brain aging. Their main goal is to increase sample size, enhance image diversity, or simulate generic disease patterns, but not to predict how an individual's brain structure will evolve over time. These methods typically do not condition on a person's baseline anatomy to preserve identity, do not explicitly enforce temporal consistency across ages, and do not focus on capturing subtle, individualized aging-related changes. Therefore, these methods cannot be directly used in our research.

### *D. Longitudinal Brain MRI Prediction*

A closer line of research to this paper is to predict future brain MRIs using one or more past scans. Early work used 2D conditional GANs to "age" MRI slices [27]. Other models, such as IdenBAT, enforce latent orthogonality to separate identity from age when simulating future MRIs, although they still rely on chronological age labels and adversarial training [11]. Other studies include perceptual–adversarial models developed to forecast rapid infant brain development at fixed intervals [9], and TR-GAN to extrapolate the next MRI from multiple past scans but only at discrete time points [10]. More recently, diffusion-based approaches have been proposed to generate cross-sectional MRIs conditioned on covariates such as age [24].

These approaches have several limitations. Some rely on 2D slice-based models, which cannot preserve the full 3D anatomy of the brain. Others require multiple prior scans to make a prediction, which limits applicability when only a single baseline MRI is available. Also, most methods predict MRIs only at predetermined time intervals and do not model aging as a continuous process. Furthermore, most of these methods depend on adversarial training, which can be unstable and may compromise anatomical fidelity, especially for subtle aging-related changes. These gaps motivate us to develop a new model, BrainPath, in this paper.

## III. Proposed Method: BrainPath

This section introduces BrainPath's architecture design, loss function, and training/inference algorithm.

**Training Phase**

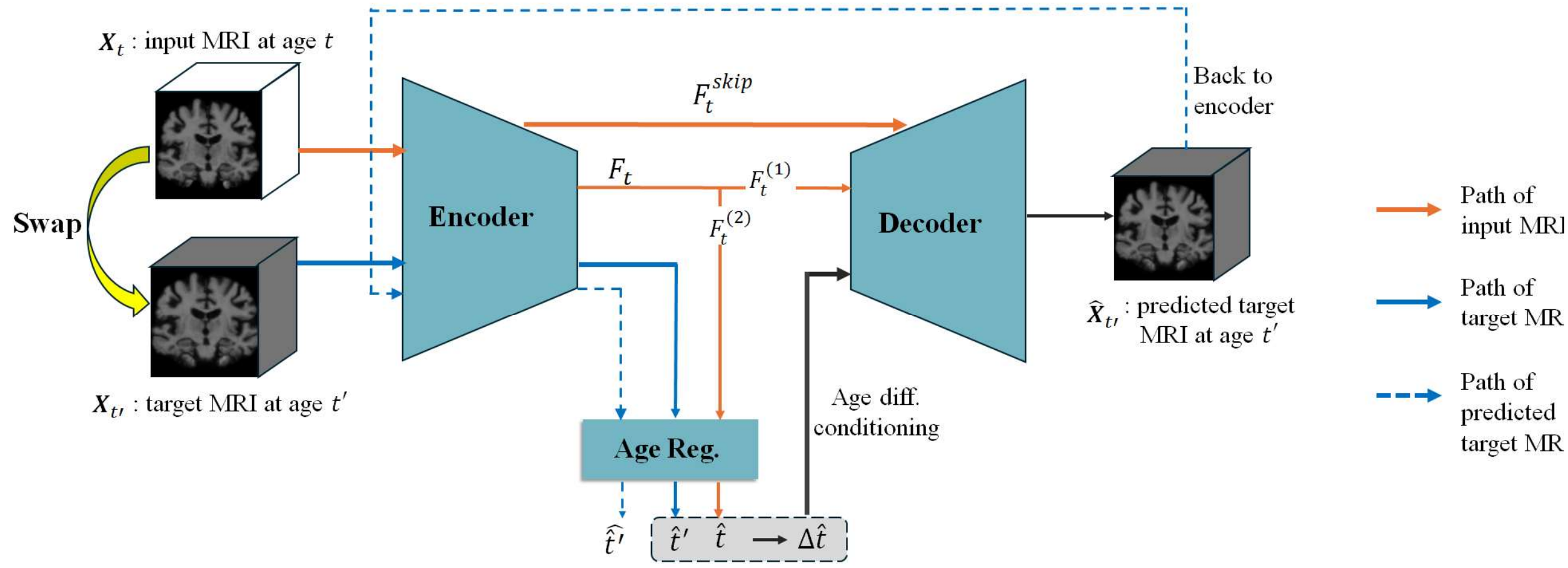

**Total loss** = Reconstruction + Age calibration + Age and structural perceptual

**Inference Phase**

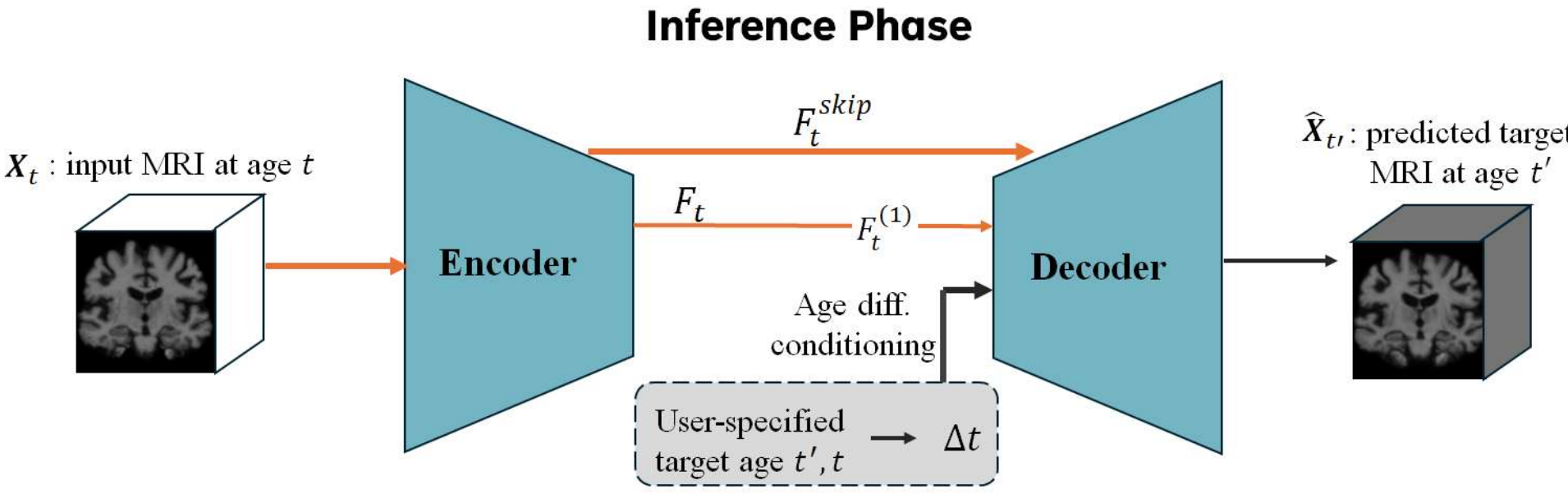

**Fig. 2. Overview of the BrainPath framework**. At the training phase (top panel), two MRIs from the same subject at age $t$ (input) and $t'$ (target) are passed through the encoder and an auxiliary age regression head, which predicts their respective brain ages, $\hat{t}$ and $\hat{t}'$. The inferred age difference $\Delta\hat{t} = \hat{t}'$-$\hat{t}$ is used as conditional signal for the decoder, together with multiscale features from the input image, to synthesize the target MRI. The training process augments the reconstruction loss with two biologically-informed losses (age calibration, age and structural perceptual) to guide learning. A swap learning strategy (yellow arrow) randomly switches input/target roles during training, promoting disentanglement between subject-specific structure and age-related variation. At inference (bottom panel), given a single MRI and a user-specified target age $t'$, BrainPath computes $\Delta t$ and synthesizes a personalized brain MRI reflecting biologically plausible aging changes.

### *A. Architecture design*

In image generation, common base architectures include GANs, diffusion models, and U-Net–based encoder–decoder frameworks. GANs are unsuitable for our setting because adversarial training is unstable, especially with high-dimensional 3D MRIs and limited sample size. This poses a high risk of losing individualized anatomical fidelity in the generated images. Diffusion models also do not fit our problem as they require extremely large training sets and long sampling processes. In contrast, U-Net–style encoder–decoders provide stable optimization, efficient use of limited data, and built-in multi-resolution skip connections that preserve fine anatomical structure.

As shown in Fig.2, BrainPath adopts a U-Net–based encoder–decoder as the backbone but introduces three key innovations: First, in the encoder, we augment the architecture with an age regression head to extract age-related representations, and further develop a novel supervision strategy grounded in biological principles to guide its learning (Sec.III.A1). Second, in the decoder, we introduce a differential age-conditioned mechanism to synthesize the aged MRI, without relying on imprecise chronological age (Sec.III. A2). Finally, we propose a swap learning strategy that implicitly disentangles subject-specific anatomy and age-dependent changes, without relying on adversarial or orthogonality constraints (Sec.III.A3).

#### *A.1 Age-aware encoder with biologically-grounded supervision*

Let $X_t$ denote the input MRI of a subject at age $t$. The encoder extracts multiscale features and passes them to the decoder. However, such a standard encoder is insufficient as it is not explicitly encouraged to learn age-related representations. To address this, BrainPath augments the encoder with an age regression head $AgeReg(.)$ and a channel-splitting mechanism. Specifically, let $F_t^{skip}$ denote the intermediate feature maps (i.e., skip connections) passed to the decoder, and let $F_t$ denote the bottleneck feature map (i.e., final encoder output). $F_t$ is split along the channel dimensions into two halves, $\left(F_t^{(1)}, F_t^{(2)}\right)$. While $F_t^{(1)}$ is passed to the decoder, $F_t^{(2)}$ is used for age prediction, i.e.,

$$\hat{t} = AgeReg(F_t^{(2)}).$$

This channel split mechanism encourages a separation between age-related and anatomy-related information, reducing interferences.

A key challenge here is to determine the appropriate supervision strategy for the age regression head. Although age regression from MRI has been widely studied (see Sec.II.A), existing methods rely on minimizing the difference between the predicted and chronological age of the subject. However, chronological age is well-known to be an imprecise proxy for biological brain age [13, 14], thus not providing adequate supervision. To address this limitation, we propose a new supervision strategy. Specifically, we leverage the fact that both the input and target MRI, $X_t$ and $X_{t'}$ are available during training. We therefore pass both images through the encoder and the age regression head to generate predictions, $\hat{t}$ and $\hat{t}'$. This design has two advantages: 1) It effectively doubles the sample size for learning aging-related representations. 2) It enables the design of a new supervision strategy that is insensitive to the inaccuracies of chronological age. Our design is based on two pieces of knowledge that are both biologically-grounded and empirically-validated [12, 28]: although an individual's brain age may not match chronological age, we know that (i) the change rates of the two types of ages are similar, and (ii) the predicted brain age is unbiased for chronological age at the population level. We translate the knowledge in (i) and (ii) into two complementary constraints in our loss function design (see Sec.II.B.1 for details).

*A.2 Differential age-conditioned Decoder*

Recall that the age regression head generates predictions, $\hat{t}$ and $\hat{t}'$, from the input and target MRI, respectively. We can compute an inferred age difference, $\Delta\hat{t} = \hat{t}' - \hat{t}$. The decoder then synthesizes the target MRI by combining this inferred age difference with multiscale features from the input image:

$$\hat{X}_{t'} = De(\Delta\hat{t}, F_t^{skip}, F_t^{(1)}).$$

The intuition is that the decoder takes the subject-specific anatomy encoded in $(F_t^{skip}, F_t^{(1)})$ and "age" it to the extent defined by $\Delta\hat{t}$. This decoder design has two clear benefits: 1) Compared to a naïve approach that conditions the generation directly on the chronological target age $t'$, our decoder will not be biased by the imprecise chronological age; 2) Compared to conditioning on the chronological age difference $\Delta t$, our decoder design allows the loss to backpropagate through the age regression head, encouraging the encoder to learn more accurate and biologically-meaningful age-related representations.

*A.3 Swap learning for implicit anatomy-aging disentanglement*

A key to successful MRI generation in our setting is to preserve subject-specific anatomical structure while accurately modeling age-related changes. This requires the model to separate, at least approximately, who the subject is (stable anatomy) from how old the brain is (aging effect). Existing methods often enforce such disentanglement explicitly using orthogonality constraints or adversarial training between identity and age representations. These approaches introduce additional loss terms and architectural complexity; also, they can oversimplify the biologically intertwined nature of anatomy and aging.

In contrast, we propose a swap learning strategy in BrainPath, which is simple yet effective. It also only encourages disentanglement implicitly, which is more realistic than orthogonality constraints. Specifically, during training, BrainPath takes two MRIs from the same subject at different ages, $(X_t, X_{t'})$, and randomly assigns one as the input and other as the target. The model uses multiscale features extracted from the chosen input image, together with the inferred age difference $\Delta\hat{t}$, to predict the target MRI. In later iterations, their roles are swapped.

The swap framework brings two desired properties to BrainPath: 1) The encoder is encouraged to learn subject-specific anatomical features that are invariant across time. 2) The decoder is encouraged to only use the inferred age difference to control aging, not confounded with static anatomical embedding. Together, these properties promote a natural disentanglement between anatomy and aging without requiring orthogonality constraints or adversarial objectives.

*B. Loss function design*

A commonly used loss function for image generation is the reconstruction loss. Using this alone is insufficient for our problem of subject-specific, high-fidelity aging MRI generation. We propose two additional losses, an age calibration loss and an age and structural perception loss.

*B.1 Age calibration loss*

A core challenge in learning age-related representations from MRI is the lack of ground-truth labels for brain age. As discussed before, chronological age is often used as a proxy, but it is known to be imprecise [13, 14]. To address this, we design an age calibration loss $\mathcal{L}_{age-cali}$ that leverages biologically-informed properties of brain aging rather than relying directly on the absolute chronological age for each individual.

Based on the knowledge that (i) the change rates of the two types of ages are similar, and (ii) the predicted brain age is unbiased for chronological age at the population level [12, 28], we propose two complementary constraints, as follows:

*(i) Temporal Consistency Constraint ($L_{diff}$)*

For a healthy subject with two MRIs at chronological ages $t$ and $t'$, we expect the change in brain age to be consistent with that in chronological age. Recall that $\hat{t}$ and $\hat{t}'$are the predicted brain ages for $X_t$ and $X_{t'}$, respectively, from BrainPath's encoder. Define

$$\Delta\hat{t} = \hat{t}' - \hat{t}, \qquad \Delta t = t' - t.$$

We enforce $\Delta\hat{t} \approx \Delta t$ by minimizing the squared difference:

$$L_{\text{diff}} = (\Delta\hat{t} - \Delta t)^2 = [(\hat{t}' - \hat{t}) - (t' - t)]^2.$$

*(ii) Population Unbiasedness Constraint ($L_{mean}$)*

Furthermore, while an individual's brain age can deviate from their chronological age, studies have suggested that brain age prediction models should be unbiased at the population level. Therefore, following the Law of Large Numbers, for a batch of N independent subjects, the expected mean of the predicted brain ages should match the mean of the chronological ages:

$$\frac{1}{N}\sum_{n=1}^{N}\hat{t}_n \approx \frac{1}{N}\sum_{n=1}^{N} t_n.$$

We encourage this property by minimizing

$$L_{\text{mean}} = \left(\frac{1}{N}\sum_{n=1}^{N}\hat{t}_n - \frac{1}{N}\sum_{n=1}^{N} t_n\right)^2.$$

Together, the age calibration loss is defined as

$$L_{\text{age-cali}} = L_{\text{diff}} + \lambda_1 L_{\text{mean}}.$$

By minimizing $L_{\text{age-cali}}$, we train BrainPath to learn brain-age representations that evolve over time in a biologicall-plausible way and remain unbiased at the group level.

*B.2 Age and structural perceptual loss*

A critical challenge in modeling longitudinal brain changes is that age-related structural changes are often subtle and localized, making them susceptible to scanning artifacts and preprocessing noise. Standard voxel-wise reconstruction losses tend to emphasize low-frequency spatial similarity, which can result in overly smooth or blurry predictions that fail to capture fine-grained aging patterns.

To address this, we introduce an age and structural perceptual loss $\mathcal{L}_{age-perc}$. The basic idea is to pass both the actual and predicted target MRIs, $X_{t'}$ and $\hat{X}_{t'}$, through the encoder, and compare their latent structural features and inferred ages. Formally, recall that in Sec. III.A1, we proposed a channel-splitting mechanism for the encoder, which splits the bottleneck feature map into two halves along the channel dimension. Under this mechanism, the bottleneck features of $X_{t'}$ and $\hat{X}_{t'}$ are split as:

$$F_{t'} = \left(F_{t'}^{(1)}, F_{t'}^{(2)}\right), \qquad \hat{F}_{t'} = \left(\hat{F}_{t'}^{(1)}, \hat{F}_{t'}^{(2)}\right).$$

respectively. The first halves indexed by '(1)' aim to capture anatomical structure, so we use a Mean Square Error (MSE) loss to encourage them to match, $MSE\left(F_{t'}^{(1)}, \hat{F}_{t'}^{(1)}\right)$. The second halves indexed by '(2)' aim to capture brain age, and are passed through the age regression head to predict $\hat{t}'$ and $\hat{\hat{t}}'$. We further encourage these predictions to match. Combining the two parts, the proposed age and structural perceptual loss is

$$\mathcal{L}_{age-perc} = MSE\left(F_{t'}^{(1)}, \hat{F}_{t'}^{(1)}\right) + \lambda_2\left(\hat{t}' - \hat{\hat{t}}'\right)^2.$$

This loss ensures that the generated MRI is not only visually similar to the target but also perceptually valid in terms of its underlying anatomical features and biological age signature.

*B.3 Total loss function*

The overall training objective is a weighted combination of the age calibration, age and structural perceptual, and reconstruction losses:

$$\mathcal{L} = \mathcal{L}_{age-cali} + \mathcal{L}_{age-perc} + \mathcal{L}_{recon},$$

where $\mathcal{L}_{recon}$ is a standard reconstruction loss to enforce voxel-wise prediction accuracy between the predicted and target MRIs, i.e.,

$$\mathcal{L}_{recon} = \lambda_3 MSE(X_{t'}, \hat{X}_{t'}).$$

*C. Algorithm*

The complete procedures for model training and inference are formalized in Algorithm 1 and Algorithm 2, respectively, where $En$ is the encoder of BrainPath.

**Algorithm 1:** BrainPath Training Procedure

**Input:** Longitudinal MRI dataset $\{(X_t)\}$
**Output**: BrainPath Model $\{En, De, AgeReg\}$.
1: Initialize $En, De, AgeReg$ with random weights
2: **while** *training* **do**
3: Sample a pair $(X_t, X_{t'})$ from same subject
4: Calculate age difference: $\Delta t \leftarrow t' - t$
5: **//Swap Learning Procedure**
6: $F_t^{(1)}, F_t^{(2)}, F_t^{skip} \leftarrow En(X_t)$
7: $\hat{t} \leftarrow AgeReg(F_t^{(2)})$
8: $F_{t'}^{(1)}, F_{t'}^{(2)}, F_{t'}^{skip} \leftarrow En(X_{t'})$
9: $\hat{t}' \leftarrow AgeReg(F_{t'}^{(2)})$
10: Predicted brain age difference: $\Delta t \leftarrow \hat{t}' - \hat{t}$
11: Predict MRI: $\hat{X}_{t'} \leftarrow De(\Delta\hat{t}, F_t^{skip}, F_t^{(1)})$
12: **//Compute Losses**
13: $\mathcal{L} \leftarrow \mathcal{L}_{age-cali} + \mathcal{L}_{age-perc} + \mathcal{L}_{recon}$
14: **Update** $En, De, AgeReg$ wieghts with $\mathcal{L}$
15: **end while**
16: **Return** BrainPath Model $\{En, De, AgeReg\}$

**Algorithm 2:** BrainPath Inference Procedure

**Input:** Base MRI $X_t$ and desired age $t'$
**Output**: $\hat{X}_{t'}$
1: Initialize $En, De, AgeReg$ with pretrained weights
2: Calculate age difference: $\Delta t \leftarrow t' - t$
3: $F_t^{(1)}, F_t^{(2)}, F_t^{skip} \leftarrow En(X_t)$
4: Predict MRI: $\hat{X}_{t'} \leftarrow De(\Delta t, F_t^{skip}, F_t^{(1)})$
5: **Return** $\hat{X}_{t'}$

## IV. Real Data Application

*A. Data Description*

We utilized two large public datasets created to support aging and AD studies, ADNI and NACC. Detailed introductions to these datasets can be found in Supplementary Material. BrainPath was trained using a randomly selected subset of 382 cognitively normal subjects from ADNI, comprising a total of 2,125 T1-weighted MRI scans. For validation, the model was tested on an independent subset of 100 ADNI subjects (203 MRI scans). To assess generalizability, a separate test set was constructed using all 582 cognitively normal subjects from NACC with an age greater than 50 years, yielding 1,933 MRI scans. Our participant and scan selection criteria included a minimum of two MRI scans per participant, all visits classified as cognitively normal, and an interval of at least two years between any two consecutive visits.

### *B. Image Preprocessing and Data Augmentation*

We followed standard well-established pipelines for image preprocessing. Specifically, T1-weighted MRI were first aligned to the MNI template with rigid transformation. Within-subject rigid registration was also performed for participants with more than one scan. The images were then conformed to 1 mm isotropic voxels with a 256 x 256 x 256 matrix, intensity normalized, and skull-stripped using FreeSurfer v7 to generate preprocessed images for the BrainPath model. The outputs are then cropped to a bounding box of $220 \times 220 \times 220$ voxels and resized to $128 \times 128 \times 128$. During training, we applied random cropping to obtain subvolumes of size $120 \times 120 \times 120$, while during inference, we used center cropping. To improve generalization, data augmentation techniques were applied including random horizontal flipping, random rotation within the range of $[-5^{\circ}, +5^{\circ}]$, and the addition of voxel-wise uniform noise sampled from $[0, 0.1]$.

### *C. BrainPath Implementation Details*

The encoder and decoder of BrainPath are three–layer ResNet blocks; the age-regression head comprises two fully–connected layers. During training, we first held out 75 subjects from the training set as a validation set to determine the optimal weighting for each loss component. Based on empirical tuning, we set $\lambda_1 = 0.1, \lambda_2 = 0.01, \lambda_3 = 500.$ The model was initially trained for 200 epochs. Then, the encoder was frozen, and we introduced the age and structural perceptual loss, training the remaining components for an additional 100 epochs. Finally, the entire dataset, including the validation subset, was used for a final round of training over 100 epochs. All experiments were conducted on four A100 or H100 GPUs using data-parallel training.

### *D. Results*

In this subsection, we present the experimental results of BrainPath, encompassing both quantitative metrics and qualitative visualizations. Our evaluation was conducted using a held-out test set from the ADNI cohort and further validated on an independent NACC dataset to assess cross-dataset generalizability.

We compared BrainPath with the most recently published model, IdenBAT [16]. Since the IdenBAT paper already demonstrated substantial improvements over earlier methods, we treat it as the state-of-the-art benchmark and do not repeat comparisons with models it has already outperformed.

#### *D.1. Quantitative and qualitative accuracy*

To evaluate accuracy of the generated MRIs, we computed three standard metrics, including structural similarity index (SSIM)[30], MSE, and Peak signal-to-noise ratio (PSNR)[31]. Additionally, we developed a specialized metric, MRI-Age-Difference MAE, to more specifically quantify the model's ability to capture age-related changes.

SSIM quantifies how well the predicted MRI preserves the anatomical structure of the true MRI by comparing local patterns of intensity, contrast, and spatial detail, with highervalues indicating closer anatomical fidelity. MSE provides a straightforward measure of error magnitude, with lower values indicating higher image similarity. For MRI-Age-Difference MAE, we trained an independent brain-age-difference predictor: this network takes two MRIs from the same subject and outputs their age difference. We then apply this predictor to each predicted MRI and its corresponding input scan, obtain the predicted age difference, and compare it to the targeted difference via MAE. This metric directly measures how accurately the synthesized MRI reflects the intended temporal change; reporting the age-difference MAE rather than absolute-age error avoids the 3–5 year error often observed in standalone age-prediction models.

Table I summarized the aforementioned metrics on both the held-out ADNI test set and the independent NACC dataset. BrainPath achieves a SSIM of 0.991, 0.0000829 MSE and with an MRI-Age-Difference MAE of 0.573 years on the ADNI test set, reflecting strong perceptual similarity between predicted and ground-truth MRIs. Compared with the ADNI held-out cohort, the independent NACC test set attains similarly high accuracy; the slight performance gap is plausibly explained by (i) the wider age span and younger age distribution in NACC, whose ranges were not observed during ADNI-based training; and (ii) residual skull tissue in a subset of NACC MRIs, which introduces additional preprocessing noise.

Also, BrainPath significantly outperforms IdenBAT in SSIM, MSE, PSNR, and MRI-Age-Difference MAE (Table I). Notably, BrainPath demonstrates substantially higher accuracy in predicting brain age difference, whereas IdenBAT is trained to generate images targeting a pre-specified chronological age, which leads to limited predictive precision in longitudinal performance.

In addition to quantitative metrics, we qualitatively demonstrated BrainPath’s generation quality in Fig. 3. Specifically, Fig. 3 present a visualization of the input MRI, the predicted future MRI, the resultant difference map, and the actual future MRI with its corresponding difference map, for two subjects selected from the ADNI and NACC datasets, respectively. In the difference maps, red regions indicate increased intensity (i.e., tissue growth), while blue regions indicate decreased intensity (i.e., tissue shrinkage). Notably, the predicted difference maps from BrainPath resemble the ground truth maps. In both examples, the most prominent changes occur near the ventricular boundaries, where brain tissue shrinkage is most evident. A global pattern of brain volume reduction is observed across the cortex, with particularly pronounced thinning at the edges. Minor discrepancies in the true difference maps, such as localized intensity increases near the outer boundary of the brain, are likely due to imperfect skull stripping or image alignment during preprocessing.

TABLE I
QUANTITATIVE METRICS FOR MRI GENERATION ACCURACY (BETTER PERFORMANCE IN EACH METRIC IS HIGHLIGHTED IN BOLD)

| Dataset | Method | SSIM(↑) | MSE(↓) | PSNR(↑) | MRI-Age-Diff MAE (↓) |
|---|---|---|---|---|---|
| ADNI (held-out) | BrainPath | **0.991** | **0.0000829** | **41.516** | **0.573** |
| | IdenBAT | 0.988 | 0.0004169 | 34.805 | 2.303 |
| NACC (independent) | BrainPath | **0.978** | **0.0003961** | **37.110** | **0.737** |
| | IdenBAT | 0.971 | 0.0013907 | 29.919 | 2.999 |

**Fig. 3.** Visualization of the results for two subjects from ADNI held out test set (left) and NACC (right). Each sub-figure shows the input MRI of the subject (first column), the true output MRI and their difference map (first row), and the predicted output MRI by BrainPath and their difference map (second row).

### *D.2. Effectiveness in capturing brain aging dynamics*

Since BrainPath is designed to predict aging trajectories in the MRI, we next evaluate whether the predicted MRIs capture aging dynamics. We want to verify that the predicted images exhibit changes consistent with the expected aging process. Biologically, during aging, the anatomical difference between two scans of the same subject, $X_{i,t'}$ and $X_{i,t}$, should increase as the absolute age difference $|t' - t|$ grows. To this end, for each subject's predicted MRIs within a ±10-year interval relative to the input MRI, we compute the mean pixel intensity difference between the predicted and input scans. The results (Fig. 4) show a monotonic increase in pixel variation with age difference, and the distribution appears approximately symmetric. This provides further evidence that BrainPath predicts realistic and temporally consistent anatomical changes that reflect expected aging dynamics.

Furthermore, to complement these quantitative results, we visualize two example aging sequences (one from ADNI and one from NACC) in Fig.5, showing input MRI and predicted MRIs at 2-year intervals, along with difference maps, to demonstrate smooth and progressive anatomical changes.

### *D.3. Preservation of subject-specific anatomy*

An important property we want to evaluate BrainPath for is if its generated images for the same subject can preserve subject identity. To assess this, we input each MRI from the test set into BrainPath and predict 12 images per subject at half-year intervals spanning -3 to +3 years. Each generated MRI is then passed through the encoder to extract the penultimate layer's output as a structural embedding. These high-dimensional embeddings are reduced to 50 dimensions using PCA, followed by projection to two dimensions via t-SNE[32]. The resulting visualization in Fig.6 shows that latent codes from the same subject cluster tightly, even across different simulated ages, indicating that BrainPath effectively preserves individual-specific anatomical characteristics during age progression.

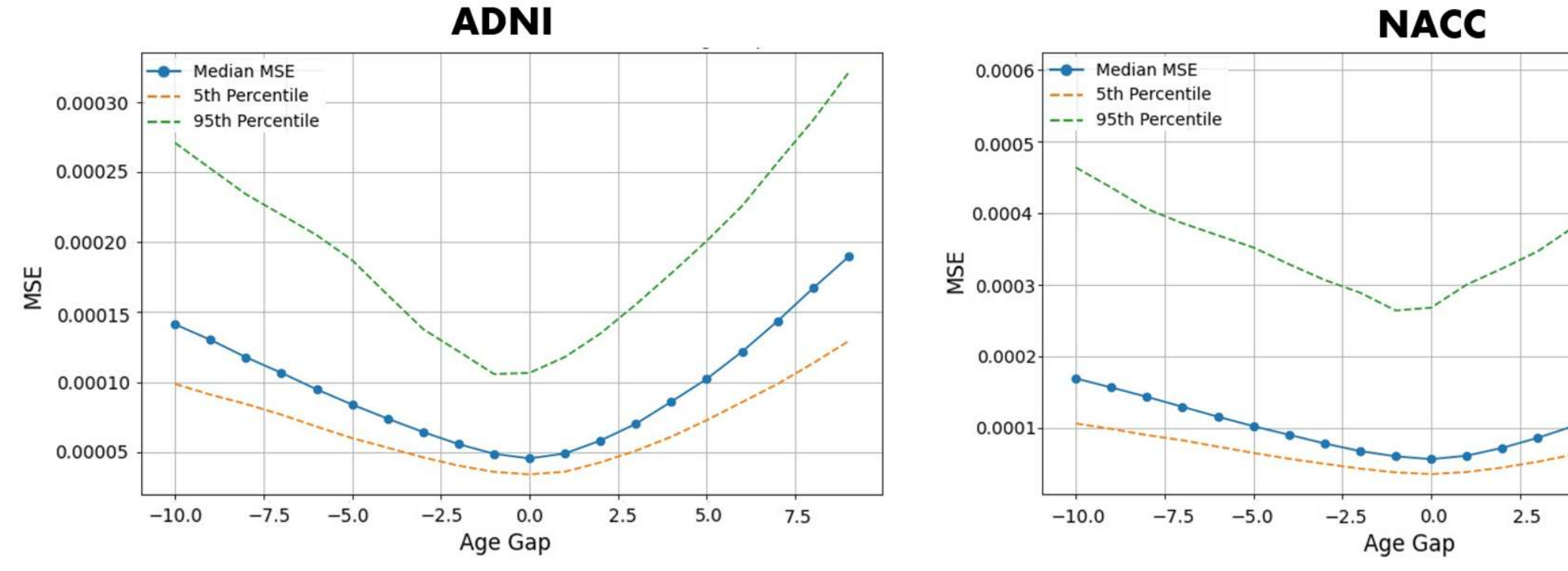

**Fig. 4.** MAE between generated and input MRI vs age difference (prediction time relative to input time) for ADNI held out test set (left) NACC (right).

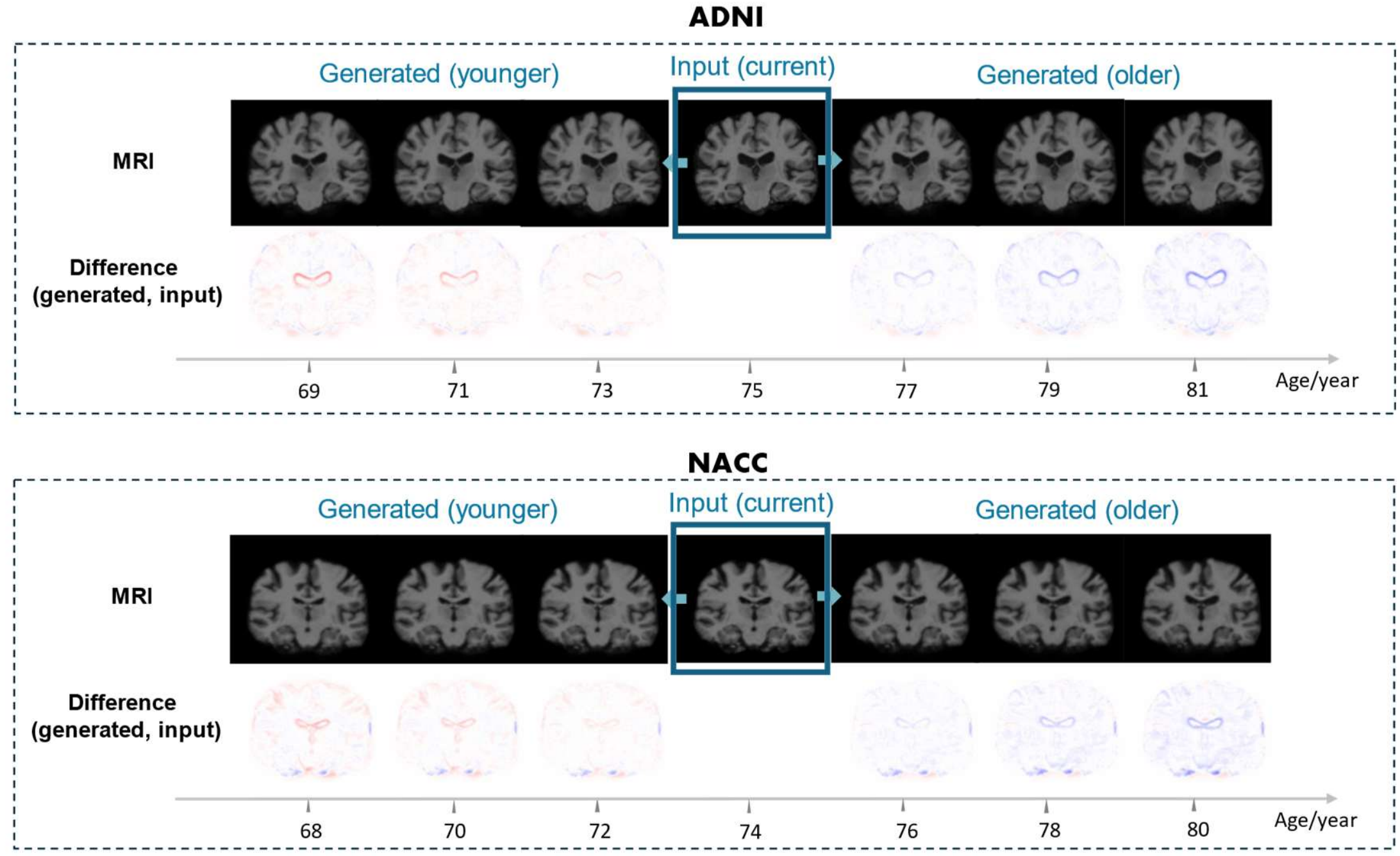

**Fig. 5.** Generated aging MRI trajectory for two subjects from ADNI held out test set (top) and NACC (bottom).

### *D .4. Ablation study*

To understand the importance of each model component in BrainPath, we conduct ablation studies by removing the swap learning (-Swap), age and structural perceptual loss (-Age & structural perceptual), or replace age calibration loss with conventionally used squared error loss (age calibration loss replaced with $\mathcal{L}_{se}$). In this context, the squared error loss refers to the standard loss function: $\mathcal{L}_{se} = (t - \hat{t})^2$ , where chronological age is directly used as the regression target of the age regression head.

We report SSIM, MRI-Age-Diff MAE and PSNR on ADNI and NACC. The results shown in Table II confirm that all components contribute to performance, with swap learning and perceptual loss playing critical roles in trajectory fidelity of the predicted MRIs.

### *D.5. Regional feature consistency*

In many studies, researchers focus on structural features of specific brain regions that are relevant to their study goals. To evaluate if the generated MRIs preserve region-specific anatomical fidelity, we use FastSurfer[33] to extract volumetric features from 31 cortical and 18 subcortical regions in both generated and true MRIs on the ADNI held-out set. We evaluate two metrics: intra-class correlation (ICC)[34] and prediction accuracy between the predicted and true volumes across subjects.

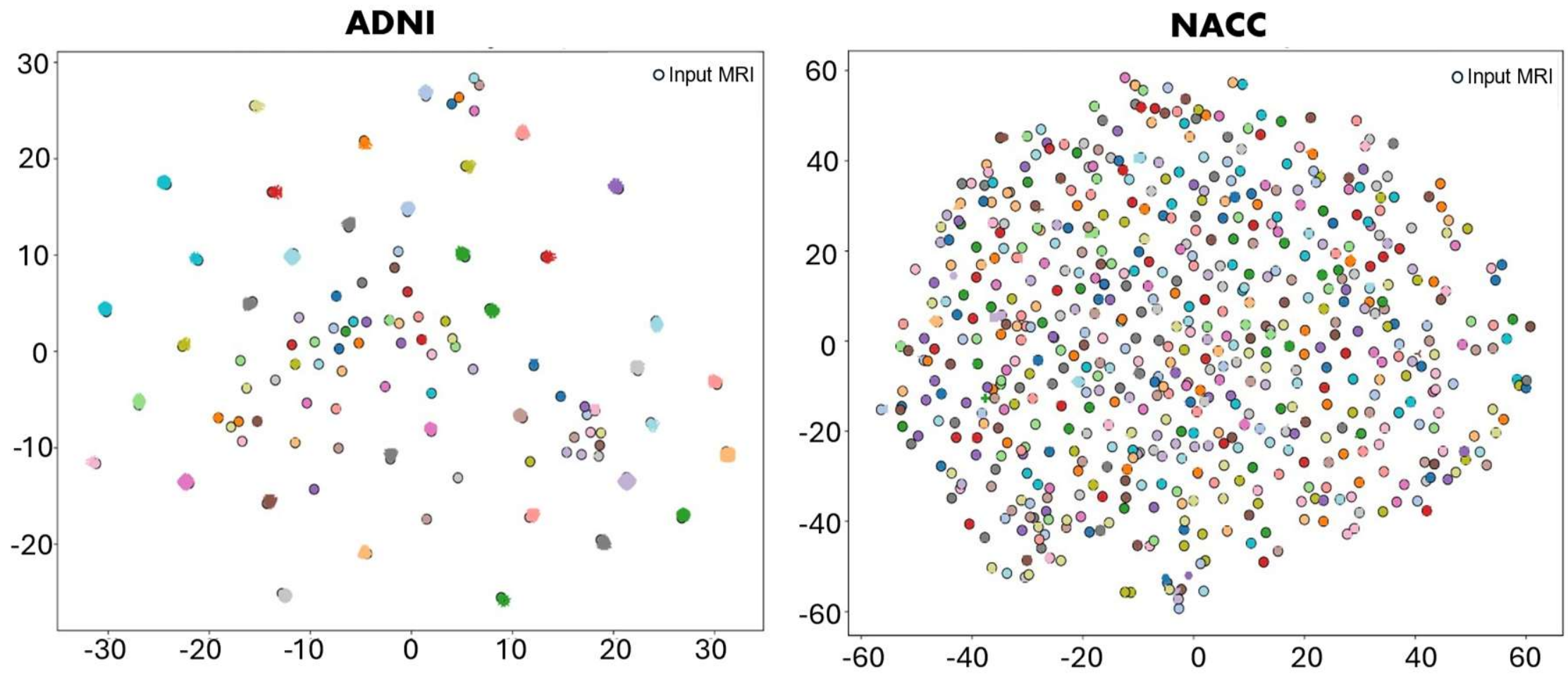

**Fig. 6.** t-SNE plot of latent features from input and generated MRIs. Points from the same subject are connected and color-coded.

TABLE II
ABLATION STUDY COMPARING MODEL VARIANTS ON ADNI AND NACC

| Model | | **BrainPath** | -Swap | -Age & structural perceptual | Age calibration loss replaced with $\mathcal{L}_{se}$ |
|---|---|---|---|---|---|
| ADNI (held-out) | SSIM(↑) | **0.99092** | 0.99071 | *0.99028* | *0.99083* |
| | MRI-Age-Diff MAE (↓) | **0.573** | 0.594 | 2.574 | 0.879 |
| | PSNR(↑) | **41.516** | 41.206 | 40.502 | 41.509 |
| NACC (independent) | SSIM(↑) | **0.97781** | 0.97731 | 0.97740 | 0.97773 |
| | MRI-Age-Diff MAE (↓) | **0.818** | 1.018 | 3.319 | 1.317 |
| | PSNR(↑) | **37.139** | 36.942 | 36.645 | 37.110 |

For each brain region, ICC was computed from paired volumetric measurements (predicted vs. ground truth) using a two-way random-effects model, defined as:

$$\text{ICC} = \frac{\sigma_{between}^2 - \sigma_{error}^2}{\sigma_{between}^2 + (k-1) \times \sigma_{error}^2}$$

where $k$ is the number of raters (here, $k = 2$: predicted and ground truth). In this formulation, $\sigma_{between}^2$ (variance between subjects) captures how much regional volumes vary across different subjects. $\sigma_{error}^2$ (residual/error variance) captures the discrepancy between predicted and true volumes within the same subject. In our context, a higher ICC indicates that the model not only approximates the absolute volumetric values but also preserves the relative ranking and inter-subject variability of brain structures, with ICC values interpreted as poor (< 0.50), moderate (0.50–0.75), good (0.75–0.90), and excellent (> 0.90). [35]. Additionally, we quantified volumetric accuracy as:

$$Accuracy = 1 - \frac{|target_{vol} - predict_{vol}|}{target_{vol}}$$

The results for the volumetric evaluation are presented in Table III, and Fig.A.1, Fig.A.2 in Supplementary Material. BrainPath achieves an average ICC of 0.9438 on ADNI and 0.9041 on NACC. For reference, FastSurfer itself reports an average ICC of 0.92 for regional volumetric features, which reflects the intrinsic reliability of this segmentation-based volume extraction method. Because our volumetric evaluation is conducted through FastSurfer, achieving ICC values that are already close to FastSurfer implies that the residual discrepancies between predicted and true volumes are smaller than the accuracy limit of the evaluation tool. In other words, the generated MRIs are of high quality such that further improvements may exceed the measurable precision of current state-of-the-art volume extraction pipelines.

## V. CONCLUSION

As populations age worldwide and aging remains a major risk factor for many diseases, there is a growing need for computational tools that can predict how an individual's brain will change over time. In this study, we proposed BrainPath, a biologically-informed 3D AI framework that, given a single structural MRI, generates synthetic longitudinal MRIs representing an individual's expected brain anatomy at arbitrary ages. Using two large public cohorts (ADNI and NACC), we demonstrated that BrainPath achieves high generation accuracy, preserves subject-specific anatomy, captures realistic aging dynamics, and generalizes across datasets, outperforming state-of-the-art baseline. These capabilities position BrainPath as a promising tool for healthcare automation, supporting early risk identification, individualized intervention planning, and sensitive detection of deviations from expected brain aging.

While our results are promising, there are some limitations of this study that point to future research. First, incorporating additional covariates such as sex, education, lifestyle, and genetics could further improve personalization and aging trajectory accuracy. Second, this work focuses on normative aging in cognitively normal individuals. Extending BrainPath to explicitly model trajectories under disease processes (e.g., MCI, AD, or other neurodegenerative disorders) would enable simulation of pathological progression and treatment effects. Third, integrating uncertainty quantification in the generation process will be essential for risk-aware clinical decision support. Finally, prospective and task-specific studies are needed to evaluate how deploying BrainPath in real-world settings can impact clinical workflows, patient outcomes, and health system resource planning.

TABLE III
ACCURACIES OF VOLUMETRIC FEATURES IN GENERATED MRIS BY BRAINPATH

| Dataset | Subcortical Accuracy | Cortical Accuracy | Overall Accuracy |
|---|---|---|---|
| ADNI (held-out) | 0.9660 | 0.9569 | 0.9627 |
| NACC (independent) | 0.9372 | 0.9503 | 0.9455 |